# Prolonging Valley Polarization Lifetime through Gate-Controlled Exciton-to-Trion Conversion in Monolayer Molybdenum Ditelluride

Qiyao Zhang,[1,2,3]† Hao Sun,[1,2,3]† Jiacheng Tang,[1,2,3] Xingcan Dai,[1] Zhen Wang,[1,2,3] Cun-Zheng Ning[1,2,3]*

*[1]Department of Electronic Engineering, Tsinghua University, Beijing 100084, China*
*[2]Frontier Science Center for Quantum Information, Beijing 100084, China*
*[3]Tsinghua International Center for Nano-Optoelectronics, Tsinghua University, Beijing 100084, China*

† These authors contributed equally to this work
\* Corresponding author. Email: cning@tsinghua.edu.cn

**Abstract:**

Monolayer 2D semiconductors provide an attractive option for valleytronics due to the valley-addressability by helicity-specific light beam. But the short valley-polarization lifetimes for excitons have hindered potential valleytronic applications. In this paper, we demonstrate a strategy for prolonging the valley polarization lifetime by converting excitons to trions through an effective gate control and by taking advantage of much longer valley polarization lifetime for trions than for excitons. In continuous-wave experiments, we found that the valley polarization for both excitons and trions increases as gate voltage is tuned away from the charge neutrality, with the degree of valley polarization increased from near zero to 38 % for excitons and to 33 % for trions. This is the first successful observation of valley-polarization in $MoTe_2$ without a magnetic field. In pump-probe experiments, we found that the intervalley scattering process of excitons is significantly suppressed as gate voltage is tuned away from charge neutrality, with the scattering time increased from 0.85 ps to ~ 2.17 ps. In contrast, the intervalley scattering rate for trions increases due to increased availability of partner charges for trion spin flipping, with the scattering time decreasing from 1.39 ns down to ~100 ps away from the charge neutrality. Interestingly, our results show that, despite the accelerated intervalley scattering, the degree of valley polarization for trions increases due to polarized trion generation from the exciton-to-trion conversion overtaking the intervalley trion scatterings. Importantly, the efficient exciton-to-trion conversion changed the dominant depolarization mechanisms from the fast electron-hole exchange for excitons to the slow spin-flip process for trions. As a result, the valley lifetime is dramatically improved by 1000 times from excitons to trions at the charge neutrality. Our results shed new light into the depolarization dynamics and the interplay of various depolarization channels for excitons and trions and provide an effective strategy for prolonging the valley polarization.

**MAIN TEXT**

**Introduction**

Valleytronics, in analogy to spintronics, aims to control and explore the valley degree of freedom (VDOF) analogously to spin degree of freedom for information processing (*1–7*). This field has been invigorated by the emergence of monolayer transition metal dichalcogenides (ML-TMDCs), which possess direct bandgaps at degenerate K/K' valleys with the locked spin orientations. Thus, optical transitions in a specific valley can be selectively excited by circularly polarized light (*5*, *8*). This permits the addressability and control of VDOF by optical means, providing an important potential for information processing stored in VDOF. However, this theoretical potential has encountered practical challenges. This is because optical transitions in TMDCs are dominated by excitons. But the valley polarization of excitons has a very short lifetime on the orders of a few picoseconds (ps) (*9*, *10*), making exciton-based valleytronics impractical. The situation is even worse for Mo-based TMDCs such as $MoTe_2$, due to the small energetic difference and the associated couplings between dark and bright excitons (*11*). Trion (T), an excitonic complex consists of exciton (X) and an extra charge, can also stably exist in TMDCs (*12*, *13*) with binding energies between 20 - 40 meVs. Compared to excitons, the lifetime of trion valley polarization has been proven to be much longer, ranging from tens to hundreds of ps (*10*, *14*, *15*). A simple comparison of valley polarization lifetimes of excitons and trions suggests a strategy of converting excitons to trions to prolong the valley polarization. Indeed, systematic gate-controlled exciton-to-trion (X-T) conversions (*16*) have been explored to realize optical gain associated with trion-electron population inversion at extremely low levels of carrier densities. Similar gate-controlled X-T conversion could enable long valley polarization so that VDOF of trions could be explored for novel information applications, such as electrically-controlled logic gate in quantum computing (*17*, *18*).

The electrical control of valley polarization has been studied in many TMDCs by performing both steady-state and time-resolved optical measurements. The electron-hole (e-h) exchange interaction, the main depolarization mechanism of excitons was shown to decrease with electric field (*19*) and with electrical gating (*20*, *21*) due to the Coulomb screening. Besides, the mutual conversion between excitons and trions, which can be controlled by electrical gating, also plays an important role in the generation and manipulation of VDOF. For instance, the X-T conversion is an important decay channel for excitons (*22*, *23*), affecting the excitons lifetime. Interestingly, it was found that the intervalley scattering of excitons can lead to fast trion depolarization due to the X-T conversion in the opposite valley in an intrinsically-doped ML-WSe$_2$ (*14*). In order to manipulate valley polarization, chemical doping (*24*) and electrical gating (*25*) were considered to effectively control the X-T conversion in continuous-wave photoluminescence (CW-PL) measurement. However, it is still unclear how the mutual conversion of X and T, especially the interplay of their depolarization processes would affect the valley polarization dynamically. Furthermore, how gate control can be employed to effectively prolong the valley polarization maintenance time? Thus, ultrafast time-resolved measurement involving X-T conversion process is necessary to explore manipulation of the VDOF. Moreover, in-depth understandings of the time scales of various dynamical processes, such as exciton/trion formation, X-T conversion, and different depolarization channels of excitons and trions, *etc.*, provide fruitful experimental evidences and shed light on the mechanisms of valley generation and manipulation.

In this paper, we systematically investigate the valley polarization and valley dynamics of both excitons and trions in monolayer molybdenum ditelluride (ML-MoTe$_2$) using CW-PL and helicity-resolved time-dependent pump-probe spectroscopy. Through the near-resonant excitation, we demonstrated the first successful observation of PL polarization without magnetic field in electrically-gated ML-MoTe$_2$ in the near-infrared wavelength range. Moreover, we show that the PL polarization and depolarization times can be effectively controlled by the gate voltage through different depolarization mechanisms. It is shown that the efficient X-T conversion changes the fast e-h exchange depolarization channel for excitons to the much slower spin-flip intervalley scattering for trions in MoTe$_2$. The generation, manipulation and depolarization mechanisms of valley polarizations in ML-MoTe$_2$ elucidated in this work enrich the understanding of the inner relationship between many-body interactions and valley dynamics, paving the way for valleytronic applications integrated on silicon platform.

Of special interest is our focus on MoTe$_2$. To date, valley polarization has been mostly observed in the visible wavelength range (*5*, *26*, *27*), while the near-infrared valley polarization (*28*) is the basic requirement for integrating valleytronic devices on Silicon platform. Monolayer molybdenum ditelluride (ML-MoTe$_2$) is an advantageous TMDC material, due to the silicon-transparent emission of excitonic transitions at room temperature, appealing to on-chip optoelectronic applications such as nanolasers (*29*), light-emitting diodes and detectors (*30*, *31*). However, due to the strong e-h exchange effect (*14*, *32*) and efficient couplings between dark/bright excitons (*11*), realization of valley polarization of excitons in ML-MoTe$_2$ has proven to be very challenging, for instance, requiring the assistance of giant magnetic field (*33–35*), which hinders the on-chip integration of valleytronic devices (*28*).

**Results**

**Gate-voltage dependence of valley polarizations for excitons and trions in CW-PL experiment**

To investigate the effect of X-T interactions on valley polarizations, we utilized an electrically-gated structure shown in Fig. 1, with Fig. 1 (A) showing the overall schematics and Fig. 1 (B) & (C) showing the cross-section view and optical microscope image of a representative device, respectively. The ML-MoTe$_2$ is mechanically exfoliated from bulk crystal and encapsulated by hexagonal boron nitride (h-BN) to enhance the material quality. The Au electrode beneath MoTe$_2$ acts as a back gate ($V_g$) and a stripe of thin graphite layer is used as a top electrode ($V_s$). (see Materials and Methods for details) The gate voltage can sensitively control the electrostatic background charges, leading to the efficient conversion to trions from the optically-excited excitons.

Although PL valley polarization and depolarization mechanisms have been extensively studied in other ML-TMDCs (*36*, *37*), the observation of valley polarization is very challenging for ML-MoTe$_2$ (*38*). The depolarization channels of excitons typically include phonon-assisted intervalley scattering and the long-range e-h exchange interactions, known as Maialle-Silva-Sham mechanism (*39*). The phonon-induced intervalley scattering mechanism has been widely discussed for MoS$_2$ and MoSe$_2$ (*27*, *37*). For this mechanism to be effective, a threshold limit for the detuning parameter, $\Delta E$, equal to twice of the longitude acoustic (LA) phonon energy $2E_{LA}$ is required to scatter both the electron and hole from K to K' valley at the same time. Here $\Delta E = \hbar\omega - E_X$, where $\hbar\omega$ is the pump laser photon energy and $E_X$ is exciton emission energy. ML-MoTe$_2$ has a much smaller LA phonon energy of $E_{LA} \sim 12$ meV than other TMDCs (*37*, *38*). This requires the pump photon energy to be very close to exciton energy and represents a more difficult experimental condition because of the linewidth broadening and efficient filtering required. This might be one of the reasons that valley polarization was not observed in ML-MoTe$_2$ previously (*38*), where a detuning parameter of 60 meV was used, much larger than $2E_{LA}$. The long-range e-h exchange interaction, which plays an important role in valley depolarization of excitons, also highly relies on the excess energy $\Delta E$. This parameter significantly affects the CW-PL degree of valley polarization (DVP). Therefore, it is crucial to study valley polarization of excitons in MoTe$_2$ system using near-resonant excitation. In our CW experiment, the detuning was chosen to be ~28 meV to minimize the depolarization mechanisms of both types mentioned above.

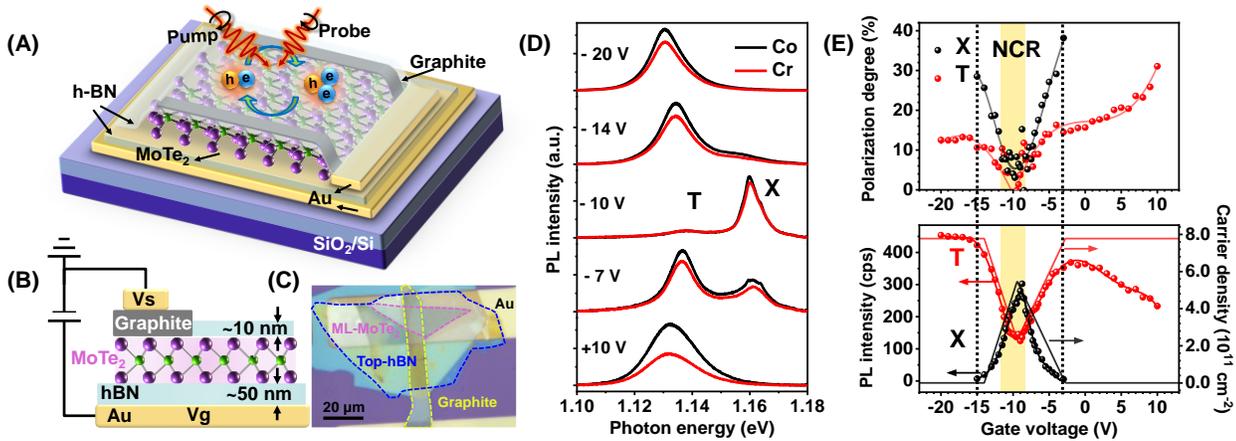

**Fig. 1. Sample structure and gate-tunable PL valley polarization.** (A) Schematic diagram of the electrically-gated ML-MoTe$_2$ device. The ML-MoTe$_2$ with top graphite electrode is encapsulated by thin layers of h-BN. And this sandwiched structure is placed on top of an Au/Ti electrode on a SiO$_2$/Si substrate. (B) Cross-section diagram of the electrically-gated structure. (C) Optical microscope image of a representative device (Device #1). ML-MoTe$_2$ is marked with purple dashed line. The yellow and blue lines mark the regions of graphite and bottom h-BN, respectively. (D) The helicity-resolved CW-PL emission of Device #1 at varying gate voltages with $\Delta E$ of 28 meV. The black and red lines denote the Co- and Cross (Cr) -circularly polarized emission with respect to the polarization direction of the pump laser. (E) The gate-dependent DVP extracted from the CW-PL measurement (top) and the gate-dependent PL emission intensity (dots) and the corresponding carrier densities (lines) of excitons and trions in ML-MoTe$_2$ (bottom). Black and red

dots (lines) represent excitons and trions, respectively. The neutrally charged region (NCR) is shaded in yellow. The black dashed lines indicate the critical voltage range within which the exciton emission can be clearly resolved.

Figure 1 (D) shows the helicity-resolved CW-PL at different gate voltages at 4 K when pumped at excess energy ΔE of ~ 28 meV from bright exciton, limited by the bandwidth of the long-pass filter used in the CW-PL measurement. We notice that bright exciton has a lower transition energy than the dark exciton with the corresponding conduction band splitting of ~ 58 meV (*38*), among the largest in TMDC materials. The excess energy is much smaller than the spin-splitting in conduction bands, meaning that only bright excitons are generated by the pump laser. Similarly only bright trions are formed with electrons or holes from the opposite valley. (see SM Section S1 for discussions of trion configuration in ML-MoTe$_2$). Although the excess energy ΔE is slightly larger than $2E_{LA}$, the small energy difference between excess energy and 2LA energy ($\Delta E - 2E_{LA}$) indicates the population of LA phonon (proportional to the phonon-assisted scattering rate), given by $\langle n \rangle = 1/(e^{2E_{LA}/(\Delta E - 2E_{LA})} - 1)$ (*27*), is negligible. Thus the phonon-assisted intervalley scattering on depolarization process can be neglected in our experiments.

The MoTe$_2$ sample (Device #1) is intrinsically negatively charged and is electrically neutral at the gate voltage ($V_g$) of - 10 V where the PL shows almost no trion emission. As shown in Fig.1 (D), the PL emission intensities for the two polarizations are almost identical at - 10 V, indicating relatively low DVP even under near-resonant excitation. This shows that the strong e-h exchange is the dominant depolarization mechanism in ML-MoTe$_2$ (see SM Section S2 for details of e-h exchange mechanism). To study the valley polarization more quantitatively, we define the value of DVP as P = (Co-Cr)/(Co+Cr), where Co and Cr (Cross) are PL emission intensities in the same or opposite circular polarization with the pump laser. The extracted DVP as a function of gate voltages are plotted in Fig. 1 (E), where we see that both excitons and trions show negligible DVP in the neutrally charged region (NCR) with voltages ranging from - 12 V to - 8 V. As the density of electrostatic background charges is increased by varying gate voltages away from NCR, the DVP of both excitons and trions increases significantly. This gate-voltage dependent behavior will be explained in more detail in connection with time-resolved DVP measurements in Fig. 3 & 4 using the rate equations. The DVP of excitons increases faster and reaches the saturation (with a maximum value of 38 % at - 3 V) earlier than that of trions. While the trion DVP continuously increases with increasing background charges, and reaches a maximum value of 33 %, which is comparable to the result of 36 % measured under a magnetic field of 29 Tesla (*33*). Our result is the first observation of polarized PL in ML-MoTe$_2$ without a magnetic field. And this type of enhancement of DVP by electrical gating in MoTe$_2$ was commonly observed in different samples in our experiments. (see SM Section S3 for PL polarization of another device-Device #3). Fig. 1 (E) (bottom) shows the gate-dependent PL intensity of excitons and trions with the estimated carrier density in both K / K' valley. The carrier density is calculated by mass action law using the electrostatic charge density derived from Fermi energy at different gate voltages (as shown in SM Section S4). In the NCR, the exciton emission is much stronger than trions. As the trion density gradually surpasses exciton density by changing gate voltages away from NCR, the DVP of both excitons and trions increased. Far away from NCR, excitons are completely converted to trions. This explains that exciton data extends to much smaller range of gate voltages in Fig. 1 (E).

To gain more insight into the mechanisms of generation and manipulation of valley polarization, the phenomenological valley-resolved rate equations as presented in full in SM Section S5 were utilized to analyze the effect of X-T conversion. The involved relaxation processes are schematically depicted in Fig. 2 (A), including intravalley and intervalley decay channels, as bordered by purple and green dashed lines. As shown in detail in SM Section S5, the DVP for excitons and trions can be expressed as

$$P_X = P_0 \frac{\Gamma_X^{ia}}{\Gamma_X^{ia} + \Gamma_X^{ir}} \tag{1a}$$

$$P_T = P_X \frac{\Gamma_T^{ia}}{\Gamma_T^{ia} + \Gamma_T^{ir}} \tag{1b}$$

where $P_0$ is the initial DVP for excitons, $\Gamma_X^{ia}$ ( $\Gamma_T^{ia}$ ) and $\Gamma_X^{ir}$ ( $\Gamma_T^{ir}$ ) denote the intravalley and intervalley decay rate of excitons (trions), respectively. The effective intravalley decay rate of excitons, $\Gamma_X^{ia}$ is given by the sum of carrier recombination $\Gamma_X^{r}$ and X-T conversion $\Gamma_{XT}$, or $\Gamma_X^{ia} = \Gamma_X^{r} + \Gamma_{XT}$. For trions, $\Gamma_T^{ia}$ represents only recombination rate $\Gamma_T^{r}$, or $\Gamma_T^{ia} = \Gamma_T^{r}$, since dissociation of trions into excitons can be neglected at low temperatures. The intervalley decay rate $\Gamma_X^{ir}$ ($\Gamma_T^{ir}$) depends on the dominant mechanisms of valley depolarization in the system, equals to twice the intervalley scattering rate $2\Gamma_X^{sk}$ ($2\Gamma_T^{sk}$). (See SM Section S5 for a detailed derivation) Whereas e-h exchange scattering for excitons is often fast, the intervalley scattering for trions is dominated by the slow spin flip process (*15*) involving the switching of an extra charge between the two valleys, as we will explain in more detail in connection with the results of the pump-probe experiments. As can be seen from the expressions (1a) and (1b), the valley polarization depends on the initially generated DVP and the ratio of intravalley decay rate $\Gamma_X^{ia}$ ($\Gamma_T^{ia}$) to the total decay rate $\Gamma_X^{ia} + \Gamma_X^{ir}$ ($\Gamma_T^{ia} + \Gamma_T^{ir}$). For excitons, the initial polarization degree $P_0$ is introduced to take into account of the unavoidable intervalley processes due to the existence of impurities or defects (*32*). So in reality, $P_0$ is lower than perfect 100 %. The low DVP in neutrally gated ML-MoTe$_2$ shown in Fig. 1 (D) & (E) also stems from the low ratio of exciton intravalley decay rate to the total decay rate, as can be seen from equation (1a). The DVP for trions has a similar expression, except that its initial value is determined through the conversion of the polarized excitons, since we assume no direct polarized trions are generated from the pump.

The X-T conversion, as an important intravalley decay channel for excitons, plays an important role in DVP for both excitons and trions. The X-T conversion contributes to the increased DVP in two ways: 1) the conversion leaves few excitons for the ultrafast intervalley scatterings (process ① in Fig. 2 (A)); 2) the converted trions will undergo a much slower intervalley scattering (process ③ in Fig. 2 (A)). As the gate voltage varies away from NCR, the increased carrier density accelerates the X-T conversion process, leading to the rapid increase of intravalley exciton decay rate (process ② in Fig. 2 (A)). This results in more excitons decay within the K valley than those scattering to K' valley, leading to the increase of DVP for excitons $P_X$ with gate voltages away from NCR. As the rapid X-T conversion process is the main formation process for trions, the initial DVP of trions is limited to $P_X$ as shown in equation (1b), in other words, the total DVP of trions is lower than that of excitons, which is consistent with our results shown in Fig. 1 (E) within the black dashed region. Moreover, as more excitons are converted to form trions, the probability of intervalley scattering via trions is increased. Since the intervalley decay rate of trions is much slower than excitons, the intervalley scattering of total populations is still greatly suppressed even when excitons are depleted. As a result, the trion DVP continuously increases even there is no exciton emission in the system due to the increased $P_X$ influenced by gating. The electrical manipulation on X-T conversions can be seen in the relative intensity change and redistribution of excitons and trions, which is shown in detail in electrical tunable CW-PL and absorption measurements in SM Section S6.

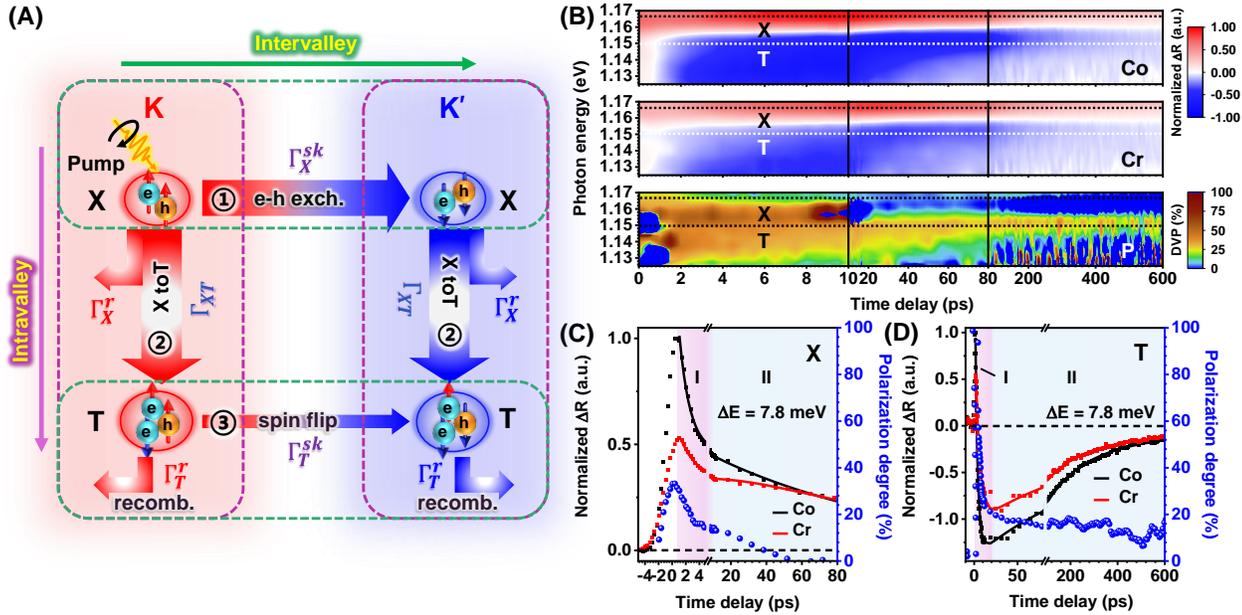

**Fig. 2. Decay processes and results of helicity-resolved pump-probe experiments.** (A) Schematic diagram of different inter- and intra-valley decay channels for excitons and trions. The colored arrows represent the flow of carrier population. After initialized by $\sigma^+$ pump, excitons in K valley face three outcomes: 1) scattered to K' valley through e-h exchange interaction (intervalley process, ①); 2) converted to trions within K valley (②) and 3) recombine radiatively or non-radiatively. Both processes 2) and 3) happen within the same valleys and do not contribute to depolarization. The trions in both valleys relax through two channels: 1) intervalley scattering to the other valley through spin flip (intervalley process, ③); and 2) radiative or non-radiative recombinations within the same valley. (B) Transient differential reflectance spectrum in colored contour map for Co- (top) and Cross- (Cr, middle) polarization with the corresponding polarization degree (P) (bottom) of Device #2 in NCR (Vg = - 13 V). The negative and positive signals in Co and Cross spectra correspond to the valley evolution dynamics of trions (T) and excitons (X), respectively. (C) & (D) Near-resonant helicity-resolved differential reflectance signals of excitons (C) and trions (D) with excess energy of 7.8 meV for Device #3 at NCR, where the right axis (blue dots) shows the DVP obtained from the differential reflectance signals. The light purple and blue shadowed areas indicate the two stages of carrier evolution dynamics.

**Helicity-resolved time-dependent pump-probe measurements**

To investigate different mechanisms of valley depolarization for excitons and trions, we performed helicity-resolved pump-probe spectroscopy on ML-MoTe$_2$ with close-to-resonance excitation at 4 K with a ~ 500 fs pump laser. (see Materials and Methods for details) The transient differential reflectance $\Delta R$, the increase of the reflection intensity of probe pulse induced by pumping, is proportional to the negative absorption, $\Delta R \propto -\alpha(0)$. Importantly, the helicity-resolved time evolution of differential reflectance data allows us to determine important time constants of intra- and inter-valley processes. To do this, we fit the experimental data by the solutions of the valley-resolved rate equations (see SM Section S5) which contain various decay and scattering rates. The intravalley decay time is fitted by an exponential function by summing up signals in both valleys, without considering the influence of intervalley scattering. And the DVP is fitted by an exponential function to obtain intervalley decay rate.

Figure 2 (B) shows the mapping of such differential reflectance in the 2D space of the delay time-probe energy with a pump energy of 1.181 eV ($\Delta E$ =15 meV) for Device #2, with the same electrically-gated structure as Device #1. The exciton and trion resonances can be identified in the broadband spectra, consistent with the absorption resonance identified by CW-PL and reflectance measurement. The peak at ~ 1.166 eV corresponds to excitons. The positive values near the exciton resonance is due to the bleaching of exciton absorption. The negative peak at ~ 1.152 eV corresponds to trions as a result of bandgap renormalization with calculated pump induced carrier density of ~ $3 \times 10^{12}$ cm$^{-2}$ (see SM Section S7). The co-polarized data (Co) reflects the decay processes of excitons and trions within the excitation valley, while the cross-polarization signal (Cr) depicts the population redistribution dynamics between K and K' valleys due to the aforementioned depolarization processes. The clear difference between the Co and Cr signals versus time delays indicates the unbalanced population evolution in K / K' valleys. The corresponding DVP mapping labelled with P in the bottom panel of Fig. 2 (B) shows remarkably different dynamics of DVPs for excitons and trions. The exciton DVP reaches its maximum value of ~ 40 % within the first few ps and quickly disappears within tens of ps, while the trion DVP almost maintains at a constant value of ~ 20 % for over 600 ps, much longer maintenance of DVP than that of excitons.

The different behaviors of DVP dynamics of excitons and trions can be further examined by varying both pump and probe energies. The detailed transient absorption results and more discussions of the carrier behaviors at different pump and probe energies are shown in SM Section S8. While our CW experiments earlier were limited to within tens of meV of excess energy, the valley-resolved pump-probe experiments can be performed at almost the exact resonances, as shown in Fig. 2 (C) & 2 (D) for excitons and trions respectively in neutrally gated Device #3. After initialized by the intense $\sigma^+$ circularly-polarized pump, both excitons and trions are populated in the K valley due to the optical selection rule. Both the exciton and trion signals rise near zero time delay due to the bleaching of absorption as the formation of excitons and trions. Then the kinetic evolution of carriers can be divided into two stages, as shaded by the light purple (I) and blue (II) region. In stage I, the transient differential reflectance signal of both excitons and trions drops fast in the first few picoseconds, and the trions signal changes into negative values. In stage II, both excitons and trions exhibit a slower decay than in stage I.

For the exciton polarization dynamics shown in Fig. 2 (C), in stage I, a fast decay of exciton population in both K and K' valleys is attributed to the two processes that happen on similar time scales: the X-T conversion and the recombination process, or $\Gamma_X^{ia} = \Gamma_X^r + \Gamma_{XT}$. We summed up the exciton populations in both valleys and fitted the resulting time-dependence in this stage to obtain an intravalley lifetime, or $1/\Gamma_X^{ia}$ ~ 1.66 ± 0.1 ps. In stage II, the slow decay process is related to the localized excitons possibly induced by deep-level defects in the sample. The extracted exciton DVP correspondingly reflects the depolarization process of free and localized excitons. The DVP reaches a maximum value of 33% and then decays in two similar stages: with the fast decay component assigned to the depolarization of free excitons, while the slow decay component stemming from the localized excitons. Since the slow process associated with the localized exciton typically occurs after the valley depolarization process of free excitons have completed, we will only focus on the time delay within stage I. The observed low value of the initial DVP in the pump-probe experiment is related to the limited DVP generated due to defects as mentioned before. And this short decay time which greatly influence the initial DVP of excitons can not be clearly resolved due to the much shorter time resolution than the pulse width of pump laser used in our experiment (~ 500 fs). Within the decay process of free excitons, the DVP can be exponentially fitted to obtain the intervalley decay rate $\Gamma_X^{ir}$, which equals to $2\Gamma_X^{sk}$. The fitted intervalley decay time of excitons ($1/\Gamma_X^{ir}$) of ~ 0.85 ps is about one half of the intravalley exciton decay time, leading to the relatively low PL DVP as can be seen from equation (1a).

As for the trion polarization dynamics shown in Fig. 2 (D), the decreasing signals for both polarization components in stage I indicates the trion formation process from excitons (or X-T conversion) (*22*), which is nearly on the same time scale as the free exciton decay time. With the $\sigma^+$ excitation, the intervalley trions in K valley are formed mainly from free excitons excited in the K valley with an electron or hole in the K' valley. And the trions in K' valley are formed in two ways: 1) X-T conversion from intervalley scattered excitons by e-h exchange to K' valley with electrons or hole from K valley (process ① and ② in Fig. 2 (A)), 2) intervalley scattering of trions by spin flipping from K valley to K' valley (process ③ in Fig. 2 (A)). The initial trion DVP, different from that of excitons, attains the maximum value approaching 100 %, due to the faster X-T conversion than intervalley formation of trions. And the initially generated intervalley excitons related to impurities and defects will convert to trions much slower than that of free excitons. Thus, as a result of rapid intervalley scattering of excitons and simultaneous conversion to trions, trions are formed rapidly from excitons in K' valley with electrons or hole from K valley. The formation of trions in both valleys leads to the rapid decay of the trion DVP, closely correlated with the X-T conversion process. After this fast decay process, the free excitons in the sample are almost completely depleted through either recombination or conversion to trions. In stage II, the trion populations start to decay, which can be well fitted by a biexponential function, corresponding to the recombination of mobile and immobile trions (*40*). In stage II, the trion DVP almost maintains at ~ 20 % for over hundreds of ps, much longer than that of excitons. This is because the flipping of the additional charge from one valley to the other requires a large extra momentum and thus is a process of low probability, leading to the slow intervalley decay rate for trions.

**Electrical manipulation of valley dynamics via X-T interactions**

To take a closer look at the valley depolarization process via X-T conversion, we systematically investigated the gate-dependent polarization dynamics for excitons and trions as shown in Fig. 3 & 4, respectively. Fig. 3 (A) displays the differential reflectance and evolution of DVP for excitons in Device #3 at three representative voltages: - 15 V, - 10 V, and 1 V, corresponding to the cases of positively charged, charge neutrality, and negatively charged, respectively. (see SM Section S9 for the details of the electrical tuning results) We focus here on the dynamics of the first 30 ps. As can be seen, the maximum DVP is quite low (under 20 %) in the case of charge neutrality (- 10 V, middle panel), while the DVP can be as high as 60 % initially and around 20 % after the first 10 ps in the charged region. The decay of DVP gets slower in the highly charged region, indicating the suppression of intervalley scattering of excitons.

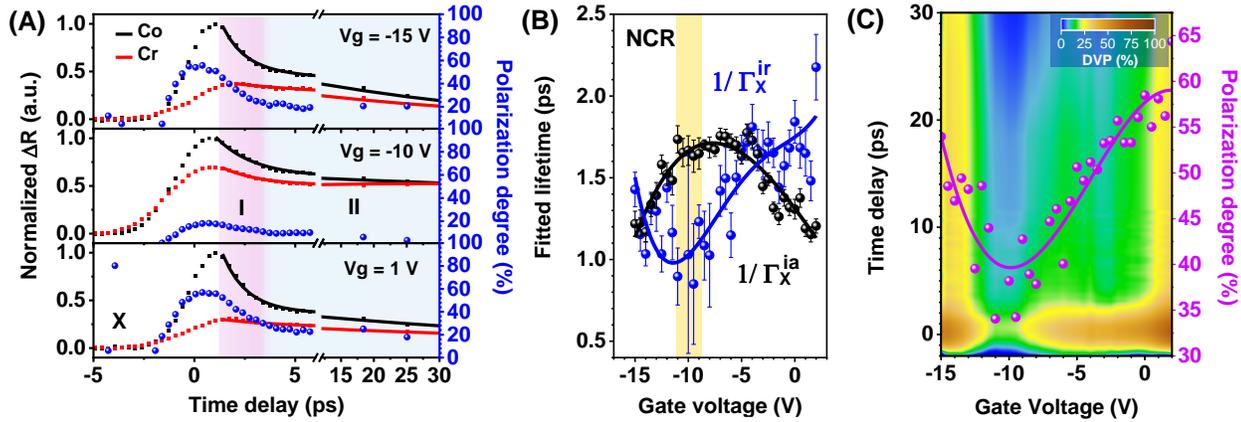

**Fig. 3. Electrical tuning of exciton valley dynamics.** (A) Gate-dependent transient differential reflectance spectra for excitons with ΔE of 7.8 meV for Device #3. The black and red dots (solid lines) represent the measured data (exponential fit) of Co and Cross (Cr) circularly-polarized signal. The signals are normalized with respect to the intensity of Co-polarized signal. The blue dots depict the dynamics of extracted DVP with scales shown on the right side. (B) The gate-dependence of fitted intravalley decay time $1/\Gamma_X^{ia}$ (black dots), intervalley decay times $1/\Gamma_X^{ir}$ (blue

dots). The solid lines are polynomial fits. (C) The colored contour of full DVP mapping in the plane of gate voltages and time delays, and the calculated CW DVP (magenta dots) for excitons using measured decay time in (B) by equation (1a) assuming $P_0$ of 100 %. The left axis is the time delay between pump and probe pulses and right axis denotes the calculated DVP.

Figure 3 (B) shows the fitted effective intravalley (intervalley) decay time $1/\Gamma_X^{ia}$ ($1/\Gamma_X^{ir}$) for excitons at different gate voltages. In the NCR around - 10 V, the intravalley decay time shows a maximum value of 1.66 ± 0.1 ps, while the intervalley decay time shows a minimum of 0.85 ± 0.37 ps. So the generated excitons are prone to intervalley scattering and the DVP at this voltage has a maximum value of only 18 %. Since the decay variation of DVP at NCR is very small, leading to a large uncertainty of the exponential fitting of polarization decay. By tuning gate voltages away from NCR, the exciton intravalley lifetime shows a slight decline, down to ~ 1.1 ps. Since the recombination rate of excitons changes negligibly with background charge density, the decrease of the effective intravalley decay time is mainly due to the acceleration of X-T conversion process, which also competes with the intervalley scattering of excitons, with both being on similar time scales.

Compared to the minor increase of intravalley lifetime, the intervalley decay time shows a much more significant increase as the gate voltage tuning away from the NCR. It more than doubles to ~ 2.17 ps in the negatively charged case around + 2 V and increases to ~ 1.43 ps in the positively charged cases at - 15V. The increase of the intervalley decay time (or the decrease of the intervalley decay rate) is the result of weakening of e-h exchange interaction due to the increasing screening from the increasing gate-generated charges as gate-voltages is tuned away from the NCR to both directions. Fig. 3 (C) shows the mapping of DVP decay with gate voltages. The maximum value of DVP increases evidently at time equals to zero when the gate voltages change away from NCR, and the maintenance time of DVP for excitons gets longer in the meantime. The DVP decay at certain gate voltages disappears rapidly, which is attributed to the fluctuations of experimental reflectance signals. The calculation results of DVP may amplify these fluctuations, but the DVP maintenance time shows an obvious trend of getting longer with gate voltages away from NCR.

This directly shows the enhancement and prolonging of DVP for excitons with gate voltages away from NCR. The detailed measurement and fitting results of intravalley decay time are shown in Fig. S4 in SM Section S5. And see the detailed helicity-resolved data at different gate voltages in SM Section S9.

The exciton DVP can also be calculated theoretically through equation (1a) by using the measured decay rates in Fig. 3 (B). Assuming the initial polarization degree $P_0$ of 100 % for simplicity, as a result, the value of $\Gamma_X^{ia}/(\Gamma_X^{ia}+\Gamma_X^{ir})$ in equation (1a) increases with gate voltages varying away from NCR, which leads to the increased DVP for excitons. The DVP is ~ 35 % at NCR, then increases to ~ 64 % at 2 V and trends toward 100 % with increased carrier density, as shown in Fig. 3 (C). However, the exciton emission decreases with charge density and is not observable at high charge densities. This theoretical calculation of exciton DVP in Device #3 is much higher than the measured value of CW results in Device #1 shown in Fig. 1 (E), but shows consistent trend of increasing DVP for excitons with gate voltages away from NCR.

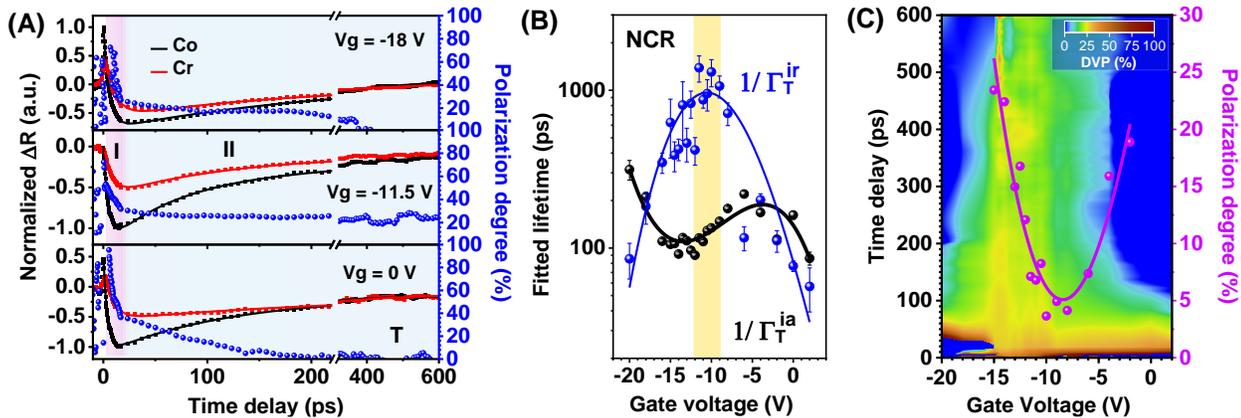

**Fig. 4. Electrical tuning of trion valley dynamics.** (A) Gate-dependent transient differential reflectance spectra of trions with ΔE of 15 meV in Device #2. The extracted DVP for trions (blue dots) can be fitted by exponential functions after the depletion of free excitons. (B) The gate-dependence of fitted intravalley lifetime $1/\Gamma_T^{ia}$ (black dots), intervalley decay time $1/\Gamma_T^{ir}$ (blue dots). The solid lines are polynomial fits, showing the data trends with gate voltages. (C) Colored contour of full DVP for trions mapping with gate voltages and time delay, and the calculated DVP (magenta dots) using measured decay time in (B) from equation (1b). The left

axis is the time delay between pump and probe pulses and right axis denotes the calculated DVP.

Similar to the case of excitons, Fig. 4 (A) shows the differential reflectance signal and the corresponding DVP for trions at different gate voltages for Device #2. The two stages of trion population evolution, *i.e.* the formation and recombination process, can be effectively manipulated by electrical gating and show clear voltage dependency. In the NCR around - 11 V, the DVP for trions maintains at a nearly constant value of ~ 20 % for over 600 ps. However, in the highly charged situations, trions lost polarization within ~ 200 ps. The detailed results of gate-dependent measurements and valley dynamics can be found in SM Section S5 and S9. Figure 4 (B) shows the extracted intravalley (recombination) lifetime $1/\Gamma_T^{ia}$ and the intervalley decay time $1/\Gamma_T^{ir}$ for trions as a function of gate voltages. These decay rates are directly related to the DVP, as indicated in equation (1b). The intravalley decay time was fitted from the time-evolution of the sum of the populations of both K and K' valleys, and the intervalley decay time can be exponentially fitted from the decay of DVP after the X-T conversion, as discussed in more detail in SM Section S5. For this device, the trion population can be well fitted using a mono-exponential function due to the high quality of the sample. In the NCR case around - 11 V, the intravalley decay time $1/\Gamma_T^{ia}$ shows a local minimum value of ~ 90 ps. Then the intravalley decay time first increases with the increased charge density then decreases as the charge density continuously increases. The increase of intravalley decay time is because the electron (hole) left by the trion recombination has more difficulties to find an unoccupied state in conduction (valence) band with the increased charge density. Then the decrease of intravalley decay time at higher negatively-charged density above - 5 V is possibly due to the enhanced carrier interactions induced by the background population.

In order to obtain the accurate intervalley decay time for trions, $1/\Gamma_T^{ir} = 1/2\Gamma_T^{sk}$, the long-term time evolution of the trion DVP was fitted to an exponential decay, as shown in SM Section S10 for the representative fitting results. The fitted intervalley decay time $\Gamma_T^{ir}$ shows a value as high as ~ 1.39 ns in the NCR around - 11 V. The intervalley decay time decreases from the NCR with increasing charge density. Fig. 4 (C) shows the DVP mapping with gate voltages and time delay for trions, depicting the maintenance of DVP is longer at NCR than in the higher charge density region. This ultralong intervalley decay time is the result of difficulty of finding a charge counterpart with an opposite spin in the case of charge neutrality. With the gate voltage deviating from NCR and increasing charge densities, it becomes easier and easier to find such charge partners. Thus the spin flipping of trions becomes easier and easier, leading to the decrease of intervalley decay time to both sides of NCR. Other spin relaxation mechanisms, including Elliot-Yafet (EY), D'yakonov-Perel' (DP), and Bir-Aronov-Pikus (BAP) mechanisms, might also contribute to the gate dependent behaviors of spin relaxation. (*5*, *41*). (See detailed discussions in SM Section S11)

It is interesting to look at the trion DVP in the CW experiments presented in Fig. 1 (E) from the perspectives of the various decay processes after we have determined the time scales of these decay rates. As shown in Fig. 4 (B), the intervalley decay time of trions $\Gamma_T^{ir}$ is approximately one order of magnitude longer than $\Gamma_T^{ia}$ in the NCR, leading to a large value of $\Gamma_T^{ia}/(\Gamma_T^{ia}+\Gamma_T^{ir})$ to be ~ 0.9 in NCR. From equation (1b), we see that $P_T = 0.9 P_X$. Moreover, the exciton DVP $P_X$ in CW-PL has been shown to be extremely low in NCR. And this low initial polarization will lead to the negligible DVP for trions in spite of the ultralong trion intervalley decay time over nanoseconds, shown in Fig. 4 (B). In the case of high charge density, the trion intervalley decay time $\Gamma_T^{ir}$ decreases to the same time scale as the intravalley decay time $\Gamma_T^{ia}$, which is not favorable for maintenance of the trion polarization. However, as the X-T conversion is enhanced, the e-h exchange of excitons is suppressed. This leads to more excitons being converted to trions before intervalley scattering. This increases the initial DVP for trion. Taking the limited exciton DVP values within the black dashed region in Fig. 1 (E) as the initial trion DVP and using the fitted intra- and inter-valley decay rate in Fig. 4 (B), we can calculate the DVP for trions using equation (1b), as shown in Fig. 4 (C). The calculated DVP for trions is ~ 5 % at - 11 V and reaches ~ 24 % at - 15 V and ~ 19 % at - 2 V. Trion DVP at higher charge density is not calculated due to the unobservable exciton DVP in CW measurements. Although the intervalley decay rate of trions becomes faster with charge density and is comparable to intravalley decay rate at high charge density, the trion polarization $P_T$ continuously increases with increasing charge density due to the increased initial polarization $P_X$ induced by the efficient X-T conversion. The electrical control on X-T conversions not only influences the exciton decay dynamics, but also plays an important role in the intervalley scattering process of excitons and trions, leading to an effective generation and control of valley polarization. Moreover, the DVP maintenance time is also dramatically improved from a few ps of excitons to hundreds of ps of trions. The ultralong maintenance time of trion DVP reflects the effective time scale for practical utilization, showing a great promise to use trions for valleytronic applications. Our results of PL DVP and intervalley decay time is also compared with various ML-TMDCs in other reports (see SM Section S12 for a summary).

**Discussions and Concluding Remarks**

We showed in this paper that, through near-resonant excitation, the DVP can be controllably increased via gate voltage from near zero to 38 % and 33 % for excitons and trions, respectively, in ML-MoTe$_2$ at 4 K. Our results show that resonant excitation is crucial to observe the valley polarization, especially for those materials that have small LA phonon energies. This explains why the PL polarization was not observed in MoTe$_2$ previously (*38*). Our first observation of the valley polarization in ML-MoTe$_2$ without a magnetic field showed that the maximum value of DVP of trions is comparable to the value obtained under a magnetic field of 29 Tesla previously (*33*). In addition, controllable valley polarization in the near-infrared wavelengths in MoTe$_2$ could be important for on-chip integrated valleytronic applications due to the low absorption of the exciton and trion emissions by Si, given the predominating roles played by Si in the current electronic and spintronic applications. Moreover, the electrical manipulation of the X-T interactions enhances the observed PL polarization. By increasing charge density, the X-T conversion is enhanced significantly, thereby suppressing the main depolarization mechanism for excitons and promoting the DVP effectively for both excitons and trions.

To elucidate how many-body interactions affect the mechanisms of valley depolarization, we systematically investigated the gate-dependent valley dynamics of both excitons and trions. The rapid X-T conversion preserve and prolong the valley polarization by converting the fast depolarization mechanism of excitons through e-h exchange interaction to the much slower spin-flip of trions, especially in the situation of the charge neutrality. As a result, an ultra-long valley polarization maintenance time exceeding 600 ps was observed for trions. This should be extremely beneficial for the detection and manipulation of valley-related effects such as valley Hall effect (*42*), and for other valleytronic applications. The valley lifetime can be further prolonged by using

TMDC heterostructures, for example, by utilizing the large separation of wave functions of electrons and holes in interlayer excitons. By integrating with chiral metasurfaces or photonic structures, the valley index can be efficiently controlled, detected, and manipulated on a long time scale. Understanding of many-body effects on valley and spin polarizations may also stimulate further research on related issues in condensed matter physics of 2D materials such as twistronics and related applications. Our results will enrich our understanding of valley dynamics in ML-TMDCs and may contribute to future on-chip valleytronic applications using $MoTe_2$.

**Materials and Methods**

**1. Electrically-gated ML-$MoTe_2$ devices fabrication**

$MoTe_2$ MLs, hexagonal boron nitride (h-BN), and graphite films were mechanically exfoliated from commercial bulk crystals (2D semiconductors or HQ graphene Inc.). The exfoliated flakes were transferred onto polydimethylsiloxane (PDMS) stamps by dry transfer method. The layer thickness is identified by atomic force microscopy and contrast of optical microscope images. The electrodes of gated structures were predefined by photolithography on a Si substrate with 300 nm $SiO_2$ and then deposited with 50/30 nm Au/Ti by electron beam evaporation. After fabrication of the electrodes, a thin h-BN film of ~ 50 nm thickness, ML-$MoTe_2$ were subsequently transferred onto the back gate electrode by the employment of micromanipulators for precise alignment. A graphite stripe of ~ 10 nm thickness were then transferred as top contacts bridging between $MoTe_2$ and the other Au/Ti electrode. A second h-BN film of ~ 10 nm thickness was transferred on top of the device for protection from contamination. All the transfer processes were carried out with the aid of PDMS as carrier stamp at moderate heating temperature to avoid possible sample degradation. Finally, the device was annealed at ~200 °C for 3 hours to reduce the air gaps between layers.

**2. Steady-state optical spectroscopy**

Steady-state optical properties of $MoTe_2$ MLs are performed in a home-build micro-PL system at cryogenic temperature of 4 K. A 632.8 nm continuous-wave HeNe laser or tunable diode laser (980 - 1060 nm, Toptica DL 100) was used as the pumping source for PL measurement. A combination of several sharp-edge long-pass filters was used for near-resonant excitation. For CW-reflectance measurements, a stabilized tungsten halogen lamp (Thorlabs SLS201) was used as the broadband source. The laser or white light excites the sample through a 100x NIR-optimized objective with NA=0.7. The reflected signal was collected by the same objective and delivered to a spectrometer (Princeton Instruments Acton 2560i) equipped with LN-cooled InGaAs CCD for detection. The spot size of pump laser was estimated to be ~ 3 µm in diameter using the knife-edge method. Electrical gating was conducted by using a commercial source meter (Keysight 2902A) or semiconductor parameter analyzer (Keysight B1500A).

### 3. Helicity-resolved pump-probe spectroscopy

Helicity-resolved ultrafast pump-probe spectroscopy was carried out using a 1040 nm femtosecond laser (pulse width of ~ 500 fs, repetition rate of 400 kHz, Spirit from Spectra Physics) was divided into pump and probe beams. The near-resonant pump excitation was achieved through optical parameter amplifier (OPA) system with carrier density estimated to be at N ~ $10^{12}$ /cm$^2$. The probe pulse is spectrally broadened by the white-light generation setup, including a focusing lens and a sapphire crystal. Both pump and probe beams are modulated by a mechanical chopper. A linear-polarizer and λ/4 wave plate are used for circularly polarization control of pump and probe pulses with ~ 95 % circularity. Both beams are combined with a non-polarized beamsplitter and sent to a 50x objective with NA=0.4. With grating-based pulse shapers, pump and probe beams can be spectrally filtered with FWHM of ~ 3 meV. The energy difference below 10 meV between pump and probe can be well resolved by the spectrometer. The differential reflectance signal $\Delta R$ is defined as $\Delta R = R_w - R_{wo}$ where $R_{w(wo)}$ is the reflected probe intensity with (without) pump and detected by an InGaAs detector using lock-in technique.